\documentclass[]{interact}

\usepackage{epstopdf}% To incorporate .eps illustrations using PDFLaTeX, etc.
\usepackage[caption=false]{subfig}% Support for small, `sub' figures and tables

\usepackage[longnamesfirst,sort]{natbib}% Citation support using natbib.sty
\bibpunct[, ]{(}{)}{;}{a}{,}{,}% Citation support using natbib.sty
\renewcommand\bibfont{\fontsize{10}{12}\selectfont}% To set the list of references in 10 point font using natbib.sty

\newenvironment{Proof}{\noindent{\em Proof:\/}}{\hfill $\Box$\par}
\newcommand{\col}{\hbox{col}}
\newcommand{\vc}{\hbox{vec}}

\theoremstyle{plain}% Theorem-like structures provided by amsthm.sty
\newtheorem{theorem}{Theorem}
\newtheorem{Lemma}{Lemma}

\theoremstyle{definition}

\theoremstyle{Remark}
\newtheorem{Remark}{Remark}
\newtheorem{Problem}{Problem}
\newtheorem{Assumption}{Assumption}
\newenvironment{Proof1}{\noindent{\em Proof:\/}}{\hfill $\Box$\par}

\begin{document}

\articletype{ARTICLE TEMPLATE}% Specify the article type or omit as appropriate

\title{Distributed Dual Objective Control of A Flywheel Energy Storage Matrix System
Under Jointly Connected Communication Network}

\author{
\name{Haiming Liu\textsuperscript{a}, Huanli Gao\textsuperscript{a}, Shuping Guo\textsuperscript{a}, and He Cai\textsuperscript{a}\thanks{CONTACT He Cai Email: caihe@scut.edu.cn}}
\affil{\textsuperscript{a}School of Automation Science and Engineering, South China University of Technology, Guangzhou 510641,  China.}
}

\maketitle

\begin{abstract}
This paper studies the distributed dual objective control problem
of a heterogenous flywheel energy storage matrix system aiming at
simultaneous reference power tracking and state-of-energy balancing.
We first prove that the solution to this problem exists by showing the
existence of a common state-of-energy trajectory for all the flywheel systems
on which the dual control objectives can be achieved simultaneously. Next, based on this
common state-of-energy trajectory, the distributed dual objective control
problem is converted into a double layer distributed tracking problem, which
is then solved by the adaptive distributed observer approach. Simulation results are
provided to validate the effectiveness of the proposed control scheme.
\end{abstract}


\section{Introduction}

Flywheel energy storage system (FESS) is an important type of energy storage system which
is valuable to modern power system, especially microgrids, by maintaining balance between power supply and demand
\citep{1zytie17,2zytie17,11zytie17}.
FESS stores energy as kinetic energy in the rotational mass of the flywheel, which has many feasible
characteristics, such as high power and energy density, fast response, low maintenance and geographical free. Among all these characteristics, FESS does no harm to environment, which makes it a promising support for the
safe and efficient utilization of intermittent renewable energy.

So far, there have been extensive study on the control of a single FESS.
In \cite{4zytie14}, the active disturbance rejection control techniques were
adopted to improve the performance of the flywheel designed for DC microgrid applications.
\cite{zytie17} considered the wide speed range operation for a FESS. A speed-dependent extended state
observer was designed to realize global linearization, which together with an adaptive feedback control
guaranteed consistent dynamic performance within the entire available operation range.
In \cite{gltse19}, by taking advantages of model predictive control, an optimal nonlinear controller was synthesized to deal with model uncertainties and external disturbances.
On the other side, there has been less effort put into the cooperative control of a
flywheel energy storage matrix system (FESMS), which consists of a group of flywheel systems to increase
the power capacity and lower down the risk of single point failure \citep{caotsg16,laitm18,sunietcta20}.
Roughly speaking, the control schemes for the cooperation of FESMS fall into two categories.
The first one is the centralized control scheme, where the information of the entire
FESMS should be known to all the subsystems. In \cite{laitm18}, a power sharing mechanism was proposed in a way that
the flywheel systems which stored the most energy have the priority to be put into use. While, this
mechanism relies on the condition that each flywheel system knows the energy level of the entire FESMS and thus
is essentially centralized. Recently, motivated by the study on multiagent system, distributed
control has been introduced to the power society to deal with various problems \citep{magdiigss16,Vallejoigss14}.
In distributed control, the subsystems shall communicate with each other over the communication network
and no global information will be needed. As a result, the communication network
can be made sparse and thus more economic efficient in contrast to centralized control.
Based on the average consensus algorithm, \cite{caotsg16} proposed a distributed control scheme for the
power sharing of a FESMS, where the sharing criteria is selected to be the current charging and discharging
 capacity. The work of \cite{caotsg16} was later extended in \cite{sunietcta20} featuring periodic
 event-triggered and self-triggered control.

In general, there are two basic control objectives for an energy storage system.
First, the power output
of the entire energy storage system should track its reference.
Second, the energy level of each energy storage unit should be balanced to keep the maximum power capacity of the entire energy storage system.
For example, for battery energy storage systems, the state-of-charge should be balanced for all the
battery packs \citep{caihutii16,litia17}, and for a general energy storage system, the state-of-charge can be redefined as
state-of-energy (SOE), which is the ratio of the current stored energy and the energy capacity
\citep{Morstyntps15,caiacess20}.
In this paper, we further consider
the distributed dual objective control problem
of a heterogenous FESMS aiming at
simultaneous reference power tracking  and SOE balancing.
To solve this problem, first,
it is proven that there exists a common SOE trajectory for all the flywheel systems
on which the dual control objectives can be achieved simultaneously.
Next, adding this
common SOE trajectory to the command generator leads to a
cascaded augmented command generator, which lends itself to the idea of
converting the distributed dual objective control
problem into a double layer distributed tracking problem.
Finally, by using the adaptive distributed observer approach \citep{lhijrnc19,zhangijss18},
the double layer distributed tracking problem is solved by a distributed control law.
In contrast to the existing results, the main contributions of this paper
are summarized as follows.
\begin{itemize}
  \item In \cite{caotsg16,caiacess20,sunietcta20}, no specific dynamics of the energy storage units were considered.
  While, in this paper, we have considered the specific rotor dynamics of the flywheel systems with heterogenous
  inertia, friction, and energy capacity parameters. In fact, the common SOE trajectory is depicted
  by all these parameters which implicitly determine how the reference power is dispatched within the FESMS.
  \item In most of the existing results, say \cite{caotsg16,caiacess20,sunietcta20,laitm18}, the communication network is assumed to be static. In this paper, it is shown that the proposed control scheme is able to work under jointly
      connected communication network, i.e.,  the communication network can be
  disconnected for all time being as long as, from time to time, the union of these disconnected
  networks is connected. This feature endows the proposed control scheme with
  two advantages. First, it
 presents certain robustness against unreliable communication environment. Second, less information exchange
 within the FESMS is needed, and thus the communication cost can be significantly reduced.
\end{itemize}

% Note that \cite{Cai2018DistributedDO} considered
%the fundamental case of general energy storage units, and in contrast, a specific FESS is considered
%in this paper. It turns out to be much more difficult to solve the dual objective control problem for FESS
%dual to the damping of flywheels.
%First, the steady state solution subject to the dual control objective in
%\cite{Cai2018DistributedDO} is static and can be identified straightforwardly. However, for the case of FESS, whether
%the steady state solution exits or not remains unknown. Second, if the steady state solution exists, how to achieve the
%dual control objective is still challenging. The contribution of this work are twofold.
%First, we have shown that, in the presence of flywheel damping,
%the steady state solution subject to the dual control objective exists and is defined by a virtual dynamic system.
%Second, under the identical damping
%condition, we have proven that the control law proposed in \cite{Cai2018DistributedDO}
%can still solve the dual objective control problem.

The rest of this paper is organized as follows.
Notation adopted in this paper are summarized in Section \ref{secnot}.
Section \ref{secsmpf} first introduces the model of the FESMS, and then
gives a mathematical problem formulation for the dual objective control problem.
The main results of this paper are presented in Section \ref{secmain}, including
the seeking of the common SOE trajectory, the design of the distributed control law,
and the stability analysis of the closed-loop system. Section \ref{secexp} validates the
effectiveness of the proposed
control scheme by a numerical example, and the paper is closed by Section \ref{seccon}.

\section{Notation}\label{secnot}

%A graph $\mathcal{G}_s =
%(\mathcal {V}_s,\mathcal {E}_s)$ is a \textbf{subgraph} of $\mathcal{G} =
%(\mathcal {V},\mathcal {E})$ if $\mathcal {V}_s \subseteq \mathcal
%{V}$ and $\mathcal {E}_s\subseteq \mathcal {E} \cap (\mathcal
%{V}_s\times \mathcal {V}_s)$.

%The edge $(i,j)$ is called an undirected edge  if $(i,j)\in
%\mathcal {E}\Leftrightarrow(j,i)\in \mathcal {E}$. The graph is called an
%undirected graph if every edge in $\mathcal {E}$ is undirected. The graph is called strongly
%connected if there exists a path between any two distinct nodes. An
%undirected and strongly connected graph is called
%connected.

%, and $a_{ij}=a_{ji}$ if
%the edge $(i,j)$ is undirected

$\mathbb{R}$ denotes the set of real numbers.
For $x_i\in \mathbb{R}^{n_i}$, $i=1,\ldots,m$, $\col(x_1,\ldots,x_m)=[x_1^T,\ldots,x_m^T]^T$. $1_n=\col(1,\ldots,1)\in \mathbb{R}^n$. For a matrix $A\in \mathbb{R}^{m\times n}$,
$\vc(A)=\col(A_1,\dots,A_n)$ where $A_i$ is the $i$th column of $A$.
$||x||$ denotes the Euclidean norm of a vector $x\in \mathbb{R}^n$ and $||A||$ denotes the
Euclidean norm of a matrix $A\in \mathbb{R}^{m\times n}$.
For a function $f(t):[0,+\infty)\rightarrow \mathbb{R}^{m\times n}$, if there
exists a nonnegative integer $q$ and $\gamma_q,\dots,\gamma_1,\gamma_0\in \mathbb{R}$ such that $||f(t)||\leq \gamma_q t^q+\cdots+\gamma_1 t+\gamma_0$ for all $t\geq 0$, then we say $f(t)$ is bounded by a
polynomial function. $\otimes$ denotes the Kronecker product of matrices.

A graph $\mathcal {G}=(\mathcal {V},\mathcal
{E})$ consists of a node set  $\mathcal {V}=\{1,\dots,N\}$
and an edge set $\mathcal {E}\subseteq \mathcal {V}\times \mathcal
{V}$. For $i, j = 1, 2, \dots, N$, $i \neq j$, an edge of $\mathcal {E}$ from node $i$ to node $j$ is denoted
by $(i,j)$, and the node $i$ is called a
neighbor of the node $j$.
If the graph $\mathcal{G}$ contains a sequence of edges of the
form $({i_1}, {i_2}), ({i_2}, {i_3}), \dots, ({i_{k}}, {i_{k+1}})$,
then the set $\{({i_1}, {i_2}), ({i_2}, {i_3}), \dots, ({i_{k}},
{i_{k+1}}) \}$ is called a directed path of $\mathcal{G}$ from node ${i_1}$
to node ${i_{k+1}}$ and  node ${i_{k+1}}$ is said to be reachable
from node ${i_{1}}$.
The graph $\mathcal {G}$ is said to contain a spanning tree if there exists a node in
$\mathcal {G}$ such that it is reachable from all the other nodes.
Given a set of $r$ graphs $\mathcal{G}_k=(\mathcal {V},\mathcal
{E}_k),~k=1,\dots,r$, the graph $\mathcal{G} =
(\mathcal {V},\mathcal {E})$ where $\mathcal {E} =\bigcup_{k=1}^r
\mathcal {E}_k$ is called the union of $\mathcal{G}_k$ and is
denoted by $\bigcup_{k=1}^r \mathcal{G}_k$.
A matrix $\mathcal{A}=[a_{ij}]\in \mathbb{R}^{N\times N}$ is said to be a weighted
adjacency matrix of a graph $\mathcal{G}$ if $a_{ii}=0$, $a_{ij}>0\Leftrightarrow(j,i)\in \mathcal {E}$,
$a_{ij}=0\Leftrightarrow(j,i)\notin \mathcal {E}$. Let $\mathcal{L}=[l_{ij}]\in
\mathbb{R}^{N\times N}$ be such that $l_{ii}=\sum_{j=1}^Na_{ij}$ and
$l_{ij}=-a_{ij}$ if $i\neq j$. Then $\mathcal {L}$ is called the
Laplacian of the graph $\mathcal{G}$.
A time signal
$\sigma(t): [0,+\infty)\rightarrow\mathcal{P}$ where $\mathcal {P} =
\{1,\dots, n\}$ for some positive integer $n$ is said to be a piecewise constant switching signal
with dwell time $\tau$ if there exists a sequence $\{t_i:i=0,1,2,...\}$ satisfying, for all
$i\geq1$, $t_{i}-t_{i-1}\geq\tau$ for some positive constant $\tau$
such that,  over each interval $[t_i, t_{i+1})$, $\sigma (t) = p$
for some integer $ p\in \mathcal{P}$. $t_0,t_1,t_2,\dots$ are called
switching instants. Given a node set
$\mathcal {V}=\{1,\dots,N\}$ and a piecewise constant switching signal $\sigma(t)$, we can define a switching graph $\mathcal{G}_{\sigma(t)}=(\mathcal {V}, \mathcal {E}_{\sigma (t)})$ where $ \mathcal
{E}_{\sigma (t)} \subseteq \mathcal {V}\times \mathcal {V}$ for all
$t \geq 0$.
Let $\mathcal{A}_{\sigma(t)}=[a_{ij}(t)]\in \mathbb{R}^{N\times N}$ denote the weighted
adjacency matrix of $\mathcal{G}_{\sigma(t)}$, $\mathcal {L}_{\sigma(t)}$
denote the Laplacian of $\mathcal{G}_{\sigma(t)}$, and
$\mathcal{N}_i(t)$ denote the set of all the neighbors of node $i$ at time instant $t$.

%For any $t \geq 0,  s>0$, let
%$\mathcal{G}([t,t+s))=\bigcup_{t_i\in[t,t+s)}
%\mathcal{G}_{\sigma(t_i)}$. $\mathcal{G} ([t,t+s))$ is called the union
%graph of $\mathcal{G}_{\sigma(t)}$ over the time interval $[t,t+s)$. Then the switching graph
%$\mathcal{G}_{\sigma (t)}$ is said to be jointly connected
%if there exists a subsequence $\{i_k\}$ of $\{i:i=0,1,2,...\}$ with
%$t_{i_{k+1}}-t_{i_k}< \nu$ for some positive $\nu$ such that the
%union graph $\mathcal{G} ([t_{i_k},t_{i_{k+1}}))$ is connected.

\section{System modeling and problem formulation}\label{secsmpf}

\begin{figure}
\begin{center}
\scalebox{0.5}{\includegraphics[viewport=120 200 570 680]{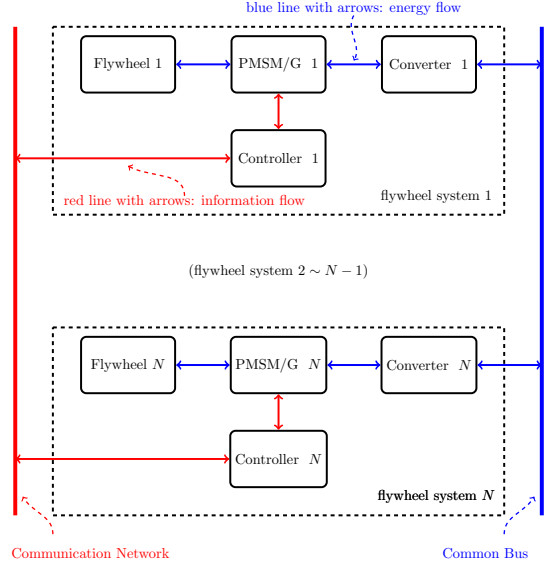}}
\caption{Configuration of the FESMS. (PMSM/G: permanent magnet synchronous motor/generator)}\label{sysconfi}
\end{center}
\end{figure}

In this paper, we consider a FESMS consisting of $N$ heterogenous flywheel systems, whose
configuration is illustrated by Fig. \ref{sysconfi}. For $i=1\dots,N$, as validated in
\cite{zytie17,gltse19}, the rotor dynamics of
the $i$th flywheel are given by:
\begin{equation}\label{sysdyn}
  I_{i}\dot{\omega}_i=-B_{vi} \omega_i+T_i
\end{equation}
where $I_{i},\omega_i,B_{vi}$ denote the inertia of the flywheel,
rotor angular velocity, and friction coefficient, respectively,
and $T_i=T_{ei}-T_{li}$ with $T_{ei}$ and $T_{li}$ denoting the electrical and mechanical load torque, respectively. The kinematic energy stored in the $i$th flywheel is $E_i=\frac{1}{2} I_{i} \omega_i^{2}$.
Let $P_i=-\dot{E}_i$. Then it follows that
\begin{equation}\label{power}
P_i=-I_{i}\omega_i\dot{\omega}_i=B_{vi} \omega_i^2-T_i\omega_i
\end{equation}
where the first part $B_{vi} \omega_i^2\triangleq P_{i,loss}$ denotes
the power loss due to friction, and the second part
$-T_i\omega_i\triangleq P_{i,out}$ denotes the net
power output of the $i$th flywheel system.
Let $\omega_{i\max}$ denote the maximum admissible angular velocity of the $i$th flywheel. Then the
energy capacity of the $i$th flywheel is given by $E_{i\max}=\frac{1}{2} I_{i} \omega_{i\max}^{2}$.
Thus, the SOE of the  $i$th flywheel is given by
\begin{equation}\label{phi}
  \phi_i=\frac{E_i}{ E_{i\max}}=\frac{\omega_i^2}{\omega_{i\max}^2}=\gamma_i\omega_i^2
\end{equation}
where $\gamma_i=1/\omega_{i\max}^2$.
By \eqref{sysdyn}, \eqref{power} and \eqref{phi}, it follows that
  \begin{equation}\label{dphi}
    \dot{\phi}_i=\frac{2\gamma_i}{I_{i}}(-B_{vi} \omega_i^2+T_i\omega_i)
    =-\frac{2B_{vi}}{I_{i}}\gamma_i\omega_i^2+\frac{2\gamma_i}{I_{i}}T_i\omega_i
    =-\frac{2B_{vi}}{I_{i}}\phi_i-\frac{2\gamma_i}{I_{i}}P_{i,out}.
  \end{equation}
In this paper, $P_{i,out}$ is taken as the control input of the $i$th flywheel system.
Let $P_{FESMS}=\sum_{i=1}^{N} P_{i,out}$ denote the power output of the entire
FESMS, and $P_{REF}(t)$ denote the reference for $P_{FESMS}$, which is assumed to be
generated by a command generator in the following form
\begin{subequations}\label{command generator}
\begin{align}
\dot{\eta}_0&=S_0\eta_0 \label{command generator.1}\\
P_{REF}&=C_0\eta_0 \label{command generator.2}
\end{align}
\end{subequations}
where $\eta_0\in \mathbb{R}^q$ is the internal state of the command generator, $S_0\in \mathbb{R}^{q\times q}$ and $C_0\in \mathbb{R}^{1\times q}$ are constant matrices.

The communication network for the FESMS is
modeled as a directed switching graph
$\mathcal{G}_{\sigma(t)}= (\mathcal{V}, \mathcal{E}_{\sigma(t)})$ , where $\mathcal{V}=\{1,\dots,N\}$ and
$\mathcal{E}_{\sigma(t)}\subseteq\{\mathcal{V} \times \mathcal{V}\}$. In here, the node $i$ of $\mathcal{V}$
is associated with the $i$th flywheel of the FESMS.
For $i,j=1,\ldots,N$, $(i,j)\in \mathcal{E}_{\sigma(t)}$
if and only if the $j$th flywheel can receive the information from the $i$th flywheel at time instant $t$.
By further taking the command generator into consideration,
we can define
an augmented switching digraph $\bar{\mathcal{G}}_{\sigma(t)}= (\bar{\mathcal{V}}, \bar{\mathcal{E}}_{\sigma(t)})$
where $\bar{\mathcal{V}}=\{0,1, \dots, N\}$ and $\bar{\mathcal{E}}_{\sigma(t)}=\mathcal{E}_{\sigma(t)}\cup\{\{0\} \times \mathcal{V}\}$. Here, the node $0$ is associated with the command generator.
For $i=1,\ldots,N$, $(0,i)\in \bar{\mathcal{E}}_{\sigma(t)}$ if and only if the $i$th flywheel can receive the information from the command generator at time instant $t$.
Let $\bar{\mathcal{A}}_{\sigma(t)}=\left[a_{i j}(t)\right] \in \mathbb{R}^{(N+1) \times(N+1)}$ be the weighted adjacency matrix of $\bar{\mathcal{G}}_{\sigma(t)}$,
$\mathcal {L}_{\sigma(t)}$
be the Laplacian of $\mathcal{G}_{\sigma(t)}$, and $H_{\sigma(t)}=\mathcal{L}_{\sigma(t)}+\hbox{diag}\left\{a_{10}(t), \ldots, a_{N 0}(t)\right\}$.
The following assumption is imposed on the communication network.

\begin{Assumption}\label{ass1}
There exists a subsequence $\{j_{k}: k=0, 1, 2, \ldots \}$ of $\{j=0,1,2,\ldots\}$
satisfying $t_{j_{k+1}}-t_{j_{k}} < \epsilon$
for some $\epsilon>0$, such that every node $i$, $i=1,\ldots,N$, is reachable from node $0$ in the union digraph
$\bigcup_{r=j_{k}}^{j_{k+1}-1} \bar{\mathcal{G}}_{\sigma(t_{r})}$.
\end{Assumption}

\begin{Remark}
  In the literature of multiagent system, Assumption \ref{ass1} is referred to as the ``jointly connected'' condition in many existing works, say \cite{lhijrnc19,shtsmcb12}.
  In contrast to the ``spanning tree'' condition for a static graph \citep{caihutii16,caiacess20}, i.e., the communication graph
  is static and contains a spanning tree with node $0$ as the root, and the ``all time connected''
  condition for a switching graph \citep{donghua16,dongzhoutie17}, i.e., the communication graph is switching and for all time being,
  it should contain a spanning tree with node $0$ as the root,
  the ``jointly connected'' condition imposes a much less restrictive requirement on the
  connectivity of the communication network. More specifically, the communication graph can be
  disconnected for all time being as long as, from time to time, the union of these disconnected
  graphs contains a spanning tree with node $0$ as the root.
  (See Fig. \ref{comnet} of Section \ref{secexp} for an example of ``jointly connected'' communication network.)
  If a control scheme can work under Assumption \ref{ass1}, then it is endowed with at least two advantages. First, it will present certain robustness against unreliable communication environment in the case that the communication link might temporarily fail due to malicious attack or
   instrument fault. Second, less information exchange among the flywheel systems and the
   command generator is needed to achieve the
  control objective, and thus the communication cost will be significantly reduced.
  \end{Remark}

Now, the dual objective control problem for the FESMS can be described as follows.

\begin{Problem}\label{prob1}
  Given systems \eqref{dphi}, \eqref{command generator} and the communication network $\bar{\mathcal{G}}_{\sigma(t)}$,
  design a distributed control law for $P_{i,out}$, such that
  \begin{equation}\label{obj1}
    \lim _{t \rightarrow \infty}\left(P_{FESMS}(t)-P_{REF}(t)\right)=0,
  \end{equation}
  and for $i, j=1, \ldots, N$, $i\neq j$,
  \begin{equation}\label{obj2}
    \lim _{t \rightarrow \infty}\left(\phi_i(t)-\phi_j(t)\right)=0,\
    \lim _{t \rightarrow \infty}\left(\dot{\phi}_i(t)-\dot{\phi}_j(t)\right)=0.
  \end{equation}
\end{Problem}
The control objective \eqref{obj1} requires that the power output of the entire FESMS shall follow its reference, and the control objective \eqref{obj2} requires SOE balancing of all the
flywheel systems.

%\begin{Remark}\label{rm1}
%In practice, the motor angular velocity $\omega_i$ of a flywheel usually has a very large feasible working region
%within $[\omega_{\min},\omega_{\max}]$ where $\omega_{\min}$ denotes the minimum admissible angular velocity.
%While, in the meantime, $\omega_i$ changes at a very slow rate due to the large inertia of the flywheel. Therefore,
%when the motor angular velocity is approaching the upper or lower limit, the flywheel can be forced offline by simply
%letting $T_i=B_{vi}\omega_i$, which cuts off the power exchange with the outside.
%In this way, we can safely presume
%that $\omega_i(t)\in [\omega_{\min},\omega_{\max}]$ for all $t\geq 0$.
%The system performance under such control strategy will be shown later in case study 5.1 of Section \ref{secexp}.
%\end{Remark}
%By Remark \ref{rm1}, the control objective \eqref{obj2} is equivalent to
% \begin{equation}\label{obj3}
%    \lim _{t \rightarrow \infty}\left(\omega_i(t)-\omega_j(t)\right)=0
%  \end{equation}
%  for $i, j=1, \ldots, N$.

\section{Main results}\label{secmain}

In this section, we first prove that the solution to Problem \ref{prob1} exists
by showing the existence of a common SOE trajectory for all the flywheel systems
on which the two control objectives \eqref{obj1} and \eqref{obj2}
in Problem \ref{prob1} can be simultaneously achieved. Next, based on this common SOE trajectory, an
augmented command generator is designed and a distributed control law is synthesized.
It is then shown that Problem \ref{prob1} can be solved by the proposed distributed control law under
Assumption \ref{ass1}.

\subsection{Problem solvability}
\begin{Lemma}\label{lem1}
If $\phi_i(t)=\psi_0(t)$ for $i=1,\dots,N$, where
\begin{equation}\label{commontrajectory}
  \dot{\psi}_0(t)=-\alpha_0\psi_0(t)-\beta_0 P_{REF}(t)
\end{equation}
with
\begin{equation*}
  \alpha_0=\frac{2\sum_{i=1}^N
  \frac{B_{vi}}{\gamma_i}}{\sum_{j=1}^N\frac{I_i}{\gamma_i}},\
  \beta_0=\frac{2}{\sum_{i=1}^N\frac{I_i}{\gamma_i}},
\end{equation*}
then it follows that
\begin{subequations}\label{ssv}
\begin{align}
P_{FESMS}(t)-P_{REF}(t)&=0 \label{ssv.1}\\
\phi_i(t)-\phi_j(t)&=0 \label{ssv.2}\\
\dot{\phi}_i(t)-\dot{\phi}_j(t)&=0. \label{ssv.3}
\end{align}
\end{subequations}
\end{Lemma}

\begin{Proof1}
Since all $\phi_i$'s are the same as \eqref{commontrajectory}, conditions \eqref{ssv.2} and \eqref{ssv.3}
are satisfied immediately. Thus, we only need to show that condition \eqref{ssv.1} is also satisfied.
For $i=1,\dots,N$, it follows that
\begin{equation}
\begin{aligned}
  \dot{\phi}_i&=-\frac{2B_{vi}}{I_i}\psi_0-\frac{2\gamma_i}{I_i}P_{i,out}
  =-\frac{2}{\sum_{i=1}^N\frac{I_i}{\gamma_i}}P_{REF}(t)-\frac{2\sum_{i=1}^N\frac{B_{vi}}
  {\gamma_i}}{\sum_{i=1}^N\frac{I_i}{\gamma_i}}\psi_0
\end{aligned}
\end{equation}
and thus
\begin{equation}
  P_{i,out}=\frac{I_i}{\gamma_i}\left(\frac{1}{\sum_{i=1}^N\frac{I_i}{\gamma_i}}P_{REF}(t)
  +\frac{\sum_{i=1}^N\frac{B_{vi}}{\gamma_i}}{\sum_{i=1}^N\frac{I_i}{\gamma_i}}\psi_0-\frac{B_{vi}}{I_i}\psi_0\right).
\end{equation}
As a result,
\begin{equation}\label{}
\begin{aligned}
 P_{FESMS}(t)&=\sum_{i=1}^NP_{i,out}(t)=\sum_{i=1}^N\frac{I_i}{\gamma_i}\left(\frac{1}{\sum_{i=1}^N\frac{I_i}{\gamma_i}}P_{REF}(t)
  +\frac{\sum_{i=1}^N\frac{B_{vi}}{\gamma_i}}{\sum_{i=1}^N\frac{I_i}{\gamma_i}}\psi_0-\frac{B_{vi}}{I_i}\psi_0\right)\\
  &=\frac{\sum_{i=1}^N\frac{I_i}{\gamma_i}}{\sum_{i=1}^N\frac{I_i}{\gamma_i}}P_{REF}(t)+
  \frac{\left(\sum_{i=1}^N\frac{I_i}{\gamma_i}\right)\left(\sum_{i=1}^N\frac{B_{vi}}{\gamma_i}\right)}{\sum_{i=1}^N\frac{I_i}{\gamma_i}}\psi_0
  -\sum_{i=1}^N\frac{B_{vi}}{\gamma_i}\psi_0\\
  &=P_{REF}(t).
\end{aligned}
\end{equation}

\end{Proof1}

\begin{Remark}
In here, an explanation is provided on how the common SOE trajectory \eqref{commontrajectory} is obtained.
  By equation \eqref{dphi} and \eqref{ssv.3}, for $i=2,\dots,N$, we have
\begin{equation}
  \frac{2\gamma_1}{I_1}(P_{1,out}+\frac{B_{v1}}{\gamma_1} \psi_0)
  =\frac{2\gamma_i}{I_i}(P_{i,out}+\frac{B_{vi}}{\gamma_i} \psi_0).
\end{equation}
Thus,
\begin{equation}
\begin{aligned}
  P_{i,out}&=\frac{\gamma_1I_i}{\gamma_iI_1}(P_{1,out}+\frac{B_{v1}}{\gamma_1} \psi_0)-\frac{B_{vi}}{\gamma_i} \psi_0\\
  &=\frac{\gamma_1I_i}{\gamma_iI_1}P_{1,out}+\left(\frac{B_{v1}I_i}{\gamma_iI_1}-\frac{B_{vi}}{\gamma_i}\right)\psi_0\\
  &=\frac{\gamma_1I_i}{\gamma_iI_1}P_{1,out}+\frac{B_{v1}I_i-B_{vi}I_1}{\gamma_iI_1}\psi_0.
\end{aligned}
\end{equation}
Therefore, by \eqref{ssv.1}, we have
\begin{equation}\label{prp1}
  \begin{aligned}
    P_{REF}(t)&=P_{1,out}+\sum_{i=2}^NP_{i,out}\\
    &=P_{1,out}+\sum_{i=2}^N\frac{\gamma_1I_i}{\gamma_iI_1}P_{1,out}+\sum_{i=2}^N\frac{B_{v1}I_i-B_{vi}I_1}{\gamma_iI_1}\psi_0\\
    &=\sum_{i=1}^N\frac{\gamma_1I_i}{\gamma_iI_1}P_{1,out}+\sum_{i=2}^N\frac{B_{v1}I_i-B_{vi}I_1}{\gamma_iI_1}\psi_0.
  \end{aligned}
\end{equation}

If the common SOE trajectory $\psi_0(t)$ exists, then for $i=1,\dots,N$, by equation \eqref{dphi}, we have
\begin{equation}
  \dot{\psi}_0=-\frac{2\gamma_i}{I_i}(P_{i,out}+\frac{B_{vi}}{\gamma_i} \psi_0).
\end{equation}
Thus, we have
\begin{equation}\label{dprp1}
  \dot{\psi}_0=-\frac{2\gamma_1}{I_1}(P_{1,out}+\frac{B_{v1}}{\gamma_1} \psi_0).
\end{equation}

Consequentially, substituting \eqref{prp1} into \eqref{dprp1} gives
\begin{equation}
  \begin{aligned}
    \dot{\psi}_0&=-\frac{2\gamma_1}{I_1}\left(\frac{P_{REF}(t)-\sum_{i=2}^N\frac{B_{v1}I_i-B_{vi}I_1}{\gamma_iI_1}\psi_0}
    {\sum_{i=1}^N\frac{\gamma_1I_i}{\gamma_iI_1}}+\frac{B_{v1}}{\gamma_1} \psi_0\right)\\
    &=-\frac{2}{\sum_{i=1}^N\frac{I_i}{\gamma_i}}P_{REF}(t)+2\left(\frac{\sum_{i=2}^N\frac{B_{v1}I_i-B_{vi}I_1}{\gamma_iI_1}}{\sum_{i=1}^N\frac{I_i}{\gamma_i}}-\frac{B_{v1}}{I_1}\right)\psi_0\\
    &=-\frac{2}{\sum_{i=1}^N\frac{I_i}{\gamma_i}}P_{REF}(t)
    +2\frac{\sum_{i=2}^N\frac{B_{v1}I_i-B_{vi}I_1}{\gamma_iI_1}-\sum_{i=1}^N\frac{B_{v1}I_i}{I_1\gamma_i}}{\sum_{i=1}^N\frac{I_i}{\gamma_i}}\psi_0\\
    &=-\frac{2}{\sum_{i=1}^N\frac{I_i}{\gamma_i}}P_{REF}(t)-\frac{2\sum_{i=1}^N\frac{B_{vi}}{\gamma_i}}{\sum_{i=1}^N\frac{I_i}{\gamma_i}}\psi_0.
  \end{aligned}
\end{equation}
\end{Remark}

\begin{Remark}
It can be seen that the common SOE trajectory \eqref{commontrajectory} is depicted by the parameters $I_i$, $\gamma_i$ and $B_{vi}$ of all the flywheel systems, and Lemma \ref{lem1} together with \eqref{dphi} implicitly reveals how the reference
power shall be dispatched within the FESMS.
\end{Remark}

\subsection{Control design}\label{seccd}
First, adding the common SOE trajectory \eqref{commontrajectory} to the command generator \eqref{command generator}
leads to the following augmented command generator:
\begin{subequations}\label{Acommand generator}
\begin{align}
\dot{\eta}_0&=S_0\eta_0 \label{Acommand generator.1} \\
P_{REF}&=C_0\eta_0 \label{Acommand generator.2} \\
\dot{\psi}_0&=-\alpha_0\psi_0-\beta_0 P_{REF}. \label{Acommand generator.3}
\end{align}
\end{subequations}

\begin{Remark}
  The augmented command generator \eqref{Acommand generator} has a cascaded structure, i.e.,
  the output $P_{REF}$ of the command generator \eqref{Acommand generator.1}-\eqref{Acommand generator.2} is the input
  of the common SOE trajectory \eqref{Acommand generator.3}. This natural cascaded structure lends itself to the
  idea of converting the distributed dual objective control problem into a double layer distributed tracking
  problem by the adaptive distributed observer approach \citep{lhijrnc19,zhangijss18}. To be more specific,
  for each flywheel system, an adaptive distributed observer, with the same cascaded double layer structure as \eqref{Acommand generator}, can be facilitated to recover the common SOE trajectory. Then, by designing a local tracking controller
  which drives the local SOE to the common SOE, the problem will be solved by Lemma \ref{lem1}.
\end{Remark}

Then, for $i=1,\dots,N$, the control law for the $i$th flywheel system is designed as
\begin{subequations}\label{ctrl1}
\begin{align}
\dot{S}_i&=\mu_S\sum_{j=0}^Na_{ij}(t)(S_j-S_i) \label{ctrl1.3}\\
\dot{C}_i&=\mu_C\sum_{j=0}^Na_{ij}(t)(C_j-C_i) \label{ctrl1.4}\\
\dot{\eta}_{i}&=S_i\eta_i+\mu_{\eta}\sum_{j=0}^{N} a_{i j}(t)\left(\eta_{j}-\eta_{i}\right) \label{ctrl1.1}\\
\hat{P}_{i,REF}&=C_i\eta_i \label{ctrl1.5}\\
\dot{\alpha}_i&=\mu_{\alpha}\sum_{j=0}^Na_{ij}(t)(\alpha_j-\alpha_i)\label{ctrl1.6}\\
\dot{\beta}_i&=\mu_{\beta}\sum_{j=0}^Na_{ij}(t)(\beta_j-\beta_i)\\
\dot{\psi}_i&=-\alpha_i\psi_i-\beta_i \hat{P}_{i,REF}+\mu_{\psi}\sum_{j=0}^Na_{ij}(t)(\psi_j-\psi_i)\label{ctrl1.7}\\
P_{i,out}&=-\frac{I_i}{2\gamma_i}\left(-\alpha_i\psi_i(t)-\beta_i \hat{P}_{i,REF}-\kappa(\phi_i-\psi_i)+\frac{2B_{vi}}{I_i}\phi_i\right)\label{ctrl1.2}
\end{align}
\end{subequations}
where $S_i\in \mathbb{R}^{q\times q}$, $C_i\in \mathbb{R}^{1\times q}$,
$\eta_i\in \mathbb{R}^q$, $\hat{P}_{i,REF},\alpha_i,\beta_i,\psi_i\in \mathbb{R}$
are the estimates of $S_0$, $C_0$, $\eta_0$, $P_{REF}$, $\alpha_0$, $\beta_0$ and
$\psi_0$, respectively.

\begin{figure}
\begin{center}
\scalebox{0.5}{\includegraphics[viewport=120 210 550 690]{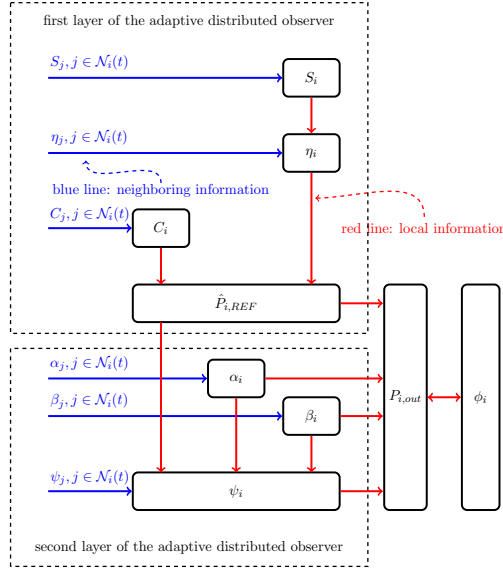}}
\caption{Block diagram of the control law \eqref{ctrl1}.}\label{controldiag}
\end{center}
\end{figure}

\begin{Remark}
  The block diagram of the control law \eqref{ctrl1} is
shown by Fig. \ref{controldiag}. Systems \eqref{ctrl1.3}-\eqref{ctrl1.5}
constitute the first layer of the adaptive distributed observer to recover $P_{REF}$,
and systems \eqref{ctrl1.6}-\eqref{ctrl1.7} constitute the second layer of the adaptive distributed observer to recover $\psi_0$. From Fig. \ref{controldiag}, it can be seen that these two layers inherit the same cascaded
structure as the augmented command generator \eqref{Acommand generator}.
\end{Remark}

\subsection{Stability analysis}\label{seccdsa}

\begin{theorem}
  Given systems \eqref{dphi} and \eqref{Acommand generator}, under Assumption \ref{ass1},
  if none of the eigenvalues of $S_0$ has positive real part, then the control law \eqref{ctrl1}
  solves Problem \ref{prob1} for any $\mu_S,\mu_C,\mu_{\eta},\mu_{\alpha},\mu_{\beta},\mu_{\psi},
  \kappa>0$.
\end{theorem}

\begin{Proof}
 For $i=1,\dots,N$, let $\bar{S}_i=S_i-S_0$, $\bar{C}_i=C_i-C_0$,
 $\bar{\alpha}_i=\alpha_i-\alpha_0$, $\bar{\beta}_i=\beta_i-\beta_0$,
 $\bar{S}=\col(\bar{S}_1,\dots,\bar{S}_N)$, $\bar{C}=\col(\bar{C}_1,\dots,\bar{C}_N)$,
 $\bar{\alpha}=\col(\bar{\alpha}_1,\dots,\bar{\alpha}_N)$,
 $\bar{\beta}=\col(\bar{\beta}_1,\dots,\bar{\beta}_N)$. Then it follows that
 \begin{subequations}
\begin{align}
\vc(\dot{\bar{S}})&=\mu_S(I_q\otimes H_{\sigma(t)}\otimes I_q)\vc(\bar{S})\\
\vc(\dot{\bar{C}})&=\mu_C(I_q\otimes H_{\sigma(t)})\vc(\bar{C})\\
\dot{\bar{\alpha}}&=\mu_{\alpha}H_{\sigma(t)}\bar{\alpha}\\
\dot{\bar{\beta}}&=\mu_{\beta}H_{\sigma(t)}\bar{\beta}.
\end{align}
\end{subequations}
 By Corollary 4 of \cite{shtsmcb12}, it follows that all
 $\vc(\bar{S}(t)),\vc(\bar{C}(t)),\bar{\alpha}(t),\bar{\beta}(t)$ will decay to zero
 exponentially as $t\rightarrow\infty$, i.e., all $\bar{S}_i(t),\bar{C}_i(t),\bar{\alpha}_i(t),\bar{\beta}_i(t)$
 will decay to zero exponentially as $t\rightarrow\infty$.
 Meanwhile, all ${S}_i(t),{C}_i(t),{\alpha}_i(t),{\beta}_i(t)$ will be bounded
 for all $t\geq 0$.

 For $i=1\dots,N$, let $\bar{\eta}_i=\eta_i-\eta_0$,
 $\bar{P}_{i,REF}=\hat{P}_{i,REF}-P_{REF}$ and $\bar{\psi}_i=\psi_i-\psi_0$.
 It follows that
 \begin{equation}\label{detai}
   \begin{aligned}
     \dot{\bar{\eta}}_i&=S_i\eta_i+\mu_{\eta}\sum_{j=0}^{N} a_{i j}(t)\left(\eta_{j}-\eta_{i}\right)-S_0\eta_0\\
     &=S_i\eta_i+S_0\eta_i-S_0\eta_i+\mu_{\eta}\sum_{j=0}^{N} a_{i j}(t)\left(\bar{\eta}_{j}-\bar{\eta}_{i}\right)-S_0\eta_0\\
     &=S_0\bar{\eta}_i+\bar{S}_i\eta_i+\mu_{\eta}\sum_{j=0}^{N} a_{i j}(t)\left(\bar{\eta}_{j}-\bar{\eta}_{i}\right)\\
     &=S_0\bar{\eta}_i+\bar{S}_i\bar{\eta}_i+\bar{S}_i\eta_0+\mu_{\eta}\sum_{j=0}^{N} a_{i j}(t)\left(\bar{\eta}_{j}-\bar{\eta}_{i}\right)
   \end{aligned}
 \end{equation}
 and
 \begin{equation}\label{dbpsi}
   \begin{aligned}
   \dot{\bar{\psi}}_i&=-\alpha_i\psi_i-\beta_i \hat{P}_{i,REF}+\mu_{\psi}\sum_{j=0}^Na_{ij}(t)(\psi_j-\psi_i)+\alpha_0\psi_0+\beta_0 P_{REF}\\
   &=-\alpha_i\psi_i+\alpha_0\psi_i-\alpha_0\psi_i-\beta_i \hat{P}_{i,REF}+\beta_i P_{REF}
   -\beta_i P_{REF}\\
   &+\mu_{\psi}\sum_{j=0}^Na_{ij}(t)(\psi_j-\psi_i)+\alpha_0\psi_0+\beta_0 P_{REF}\\
   &=-\alpha_0\bar{\psi}_i-\bar{\alpha}_i\psi_i-\bar{\beta}_i P_{REF}-\beta_i\bar{P}_{i,REF}
   +\mu_{\psi}\sum_{j=0}^Na_{ij}(t)(\bar{\psi}_j-\bar{\psi}_i)\\
   &=-\alpha_0\bar{\psi}_i-\bar{\alpha}_i\bar{\psi}_i-\bar{\alpha}_i\psi_0-\bar{\beta}_i P_{REF}-\beta_i\bar{P}_{i,REF}
   +\mu_{\psi}\sum_{j=0}^Na_{ij}(t)(\bar{\psi}_j-\bar{\psi}_i).
   \end{aligned}
 \end{equation}
 Let $\bar{\eta}=\col(\bar{\eta}_1,\dots,\bar{\eta}_N)$ and $\bar{S}_d=\hbox{block diag}\{\bar{S}_1,\dots,\bar{S}_N\}$. Then \eqref{detai} can be written into the following compact form
 \begin{equation}\label{}
   \dot{\bar{\eta}}=(I_N\otimes S_0-\mu_{\eta}(H_{\sigma(t)}\otimes I_q))\bar{\eta}+\bar{S}_d\bar{\eta}
   +\bar{S}_d(1_N\otimes \eta_0).
 \end{equation}
 By Corollary 1 of \cite{lhijrnc19}, it follows that $\lim_{t\rightarrow\infty}\bar{\eta}(t)=0$
 exponentially.
 Since none of the eigenvalues of $S_0$ has positive real part, $\eta_0(t)$ and hence $P_{REF}(t)$ are bounded by polynomial functions. As a result, $\psi_0(t)$ is also bounded by a polynomial function
 since $\alpha_0>0$. Moreover, $\eta_i=\bar{\eta}_i+\eta_0$ implies that
 $\eta_i$ is also bounded by a polynomial function since there exist
 $\rho_i,\varrho_i>0$ such that $||\bar{\eta}_i(t)||\leq \rho_ie^{-\varrho_i t}\leq \rho_i$
 for all $t\geq 0$.
 Then, noting that $\bar{P}_{i,REF}=C_i\eta_i-C_0\eta_0=\bar{C}_i\eta_i+C_0\bar{\eta}_i$
 and the fact that $\bar{C}_i$ decays to zero exponentially gives that $\bar{P}_{i,REF}$
 decays to zero exponentially.
 Let $\bar{\psi}=\col(\bar{\psi}_1,\dots,\bar{\psi}_N)$,
$\bar{\alpha}=\hbox{diag}\{\bar{\alpha}_1,\dots,\bar{\alpha}_N\}$,
$\beta=\hbox{diag}\{\beta_1,\dots,\beta_N\}$,
$\bar{\beta}=\hbox{diag}\{\bar{\beta}_1,\dots,\bar{\beta}_N\}$, and
$\bar{P}_{REF}=\col(\bar{P}_{1,REF},\dots,\bar{P}_{N,REF})$.
Then \eqref{dbpsi} can be written into the following compact form
\begin{equation}\label{}
 \dot{\bar{\psi}}=(-\alpha_0I_N-\mu_{\psi}H_{\sigma(t)})\bar{\psi}
 -\bar{\alpha}\bar{\psi}-\bar{\alpha}(1_N\otimes \psi_0)-\bar{\beta}(1_N\otimes P_{REF})-
 \beta \bar{P}_{REF}.
\end{equation}
Since  $\bar{\alpha},\bar{\beta},\bar{P}_{REF}$ decay to zero exponentially,
$\beta$ is bounded, and $\psi_0,P_{REF}$ are bounded by polynomial functions,
all $\bar{\alpha}(1_N\otimes \psi_0)$, $\bar{\beta}(1_N\otimes P_{REF})$ and
$\beta \bar{P}_{REF}$ decay to zero exponentially.
Then, again by Corollary 1 of \cite{lhijrnc19}, it follows that $\lim_{t\rightarrow\infty}\bar{\psi}(t)=0$
 exponentially. Then, similarly, $\psi_i=\bar{\psi}_i+\psi_0$
 is bounded by a polynomial function.

Substituting \eqref{ctrl1.2} into \eqref{dphi} gives
\begin{equation}\label{}
\begin{aligned}
  \dot{\phi}_i&=-\frac{2B_{vi}}{I_{i}}\phi_i-\frac{2\gamma_i}{I_{i}}\left(-\frac{I_i}
  {2\gamma_i}\left(-\alpha_i\psi_i-\beta_i \hat{P}_{i,REF}-\kappa(\phi_i-\psi_i)+\frac{2B_{vi}}{I_i}\phi_i\right)\right)\\
  &=-\alpha_i\psi_i-\beta_i \hat{P}_{i,REF}-\kappa(\phi_i-\psi_i).
\end{aligned}
\end{equation}
Let $\bar{\phi}_i=\phi_i-\psi_0$. Then we have
\begin{equation}\label{}
\begin{aligned}
  \dot{\bar{\phi}}_i&=-\alpha_i\psi_i-\beta_i \hat{P}_{i,REF}-\kappa(\phi_i-\psi_i)+\alpha_0\psi_0+\beta_0 P_{REF}\\
  &=-\kappa(\phi_i-\psi_0+\psi_0-\psi_i)-\alpha_0\bar{\psi}_i-\bar{\alpha}_i\psi_i-\bar{\beta}_i P_{REF}-\beta_i\bar{P}_{i,REF}\\
  &=-\kappa\bar{\phi}_i+\kappa\bar{\psi}_i-\alpha_0\bar{\psi}_i-\bar{\alpha}_i\psi_i-\bar{\beta}_i P_{REF}-\beta_i\bar{P}_{i,REF}.
\end{aligned}
\end{equation}
Since  $\bar{\alpha}_i,\bar{\beta}_i,\bar{P}_{i,REF},\bar{\psi}_i$ decay to zero exponentially,
$\beta_i$ is bounded, and $\psi_i,P_{REF}$ are bounded by polynomial functions,
all $\kappa\bar{\psi}_i$, $\alpha_0\bar{\psi}_i$, $\bar{\alpha}_i\psi_i$, $\bar{\beta}_i P_{REF}$, $\beta_i\bar{P}_{i,REF}$ will decay to zero exponentially.
Then, since $\kappa>0$, $\lim_{t\rightarrow\infty}\bar{\phi}_i(t)=0$, and thus
the proof is completed by invoking Lemma \ref{lem1}.

\end{Proof}

\begin{Remark}
  Assuming none of the eigenvalues of $S_0$ has positive real part merely rules out
  exponentially increasing signals, which are barely used in practice since
  the increasing rate of exponential functions is too fast.
\end{Remark}

\section{Numerical example}\label{secexp}

\begin{table}
\centering
\caption{System parameters.}
\vspace{0.5cm}
\begin{tabular}{|c|c|c|c|}
  \hline
    & $B_{vi}$ & $I_i(\mathrm{kg}\cdot \mathrm{m^2})$ & $\omega_{i,\max}(\mathrm{rad/s})$ \\ \hline
  1 & $1\times 10^{-3}$ & 0.8 & 1000 \\  \hline
  2 & $0.95\times 10^{-3}$ & 0.9 & 800 \\  \hline
  3 & $1.05\times 10^{-3}$ & 1.0 & 900 \\  \hline
  4 & $0.9\times 10^{-3}$ & 1.3 & 1200 \\
  \hline
\end{tabular}\label{table1}
\end{table}

\begin{figure}
\begin{center}
\scalebox{0.6}{\includegraphics[viewport=110 280 450 680]{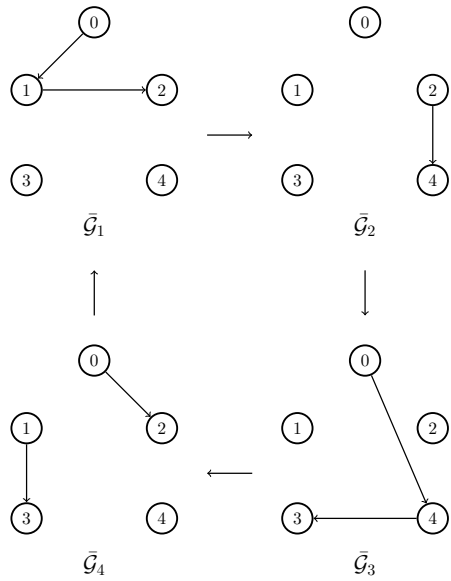}}
\caption{Communication network.}\label{comnet}
\end{center}
\end{figure}

In this section, we consider a FESMS consisting of four flywheel systems.
The communication network is shown by Fig. \ref{comnet} where $\bar{\mathcal{G}}_{\sigma(t)}$
switches among four graphs $\bar{\mathcal{G}}_1$, $\bar{\mathcal{G}}_2$, $\bar{\mathcal{G}}_3$, $\bar{\mathcal{G}}_4$ periodically every $1$s. It can be seen that all these four graphs are disconnected, while
the union of them contains a spanning tree with node $0$ as the root, and thus Assumption \ref{ass1}
is satisfied. The system parameters are given by Table \ref{table1}.
The command generator is designed as
\begin{equation}\label{}
  \dot{\eta}_0=\left(
                 \begin{array}{cc}
                   0 & 0.1 \\
                   -0.1 & 0 \\
                 \end{array}
               \right)\eta_0,\ P_{REF}=\left(
                                         \begin{array}{cc}
                                           1 & 0 \\
                                         \end{array}
                                       \right)\eta_0,\ \eta_0(0)=\left(
                                                                   \begin{array}{c}
                                                                     0 \\
                                                                     2\times 10^5 \\
                                                                   \end{array}
                                                                 \right).
\end{equation}
Thus, $P_{REF}(t)=20\sin (0.1 t)$ kw.

\begin{figure}
\begin{center}
\scalebox{0.6}{\includegraphics[viewport=40 230 530 580]{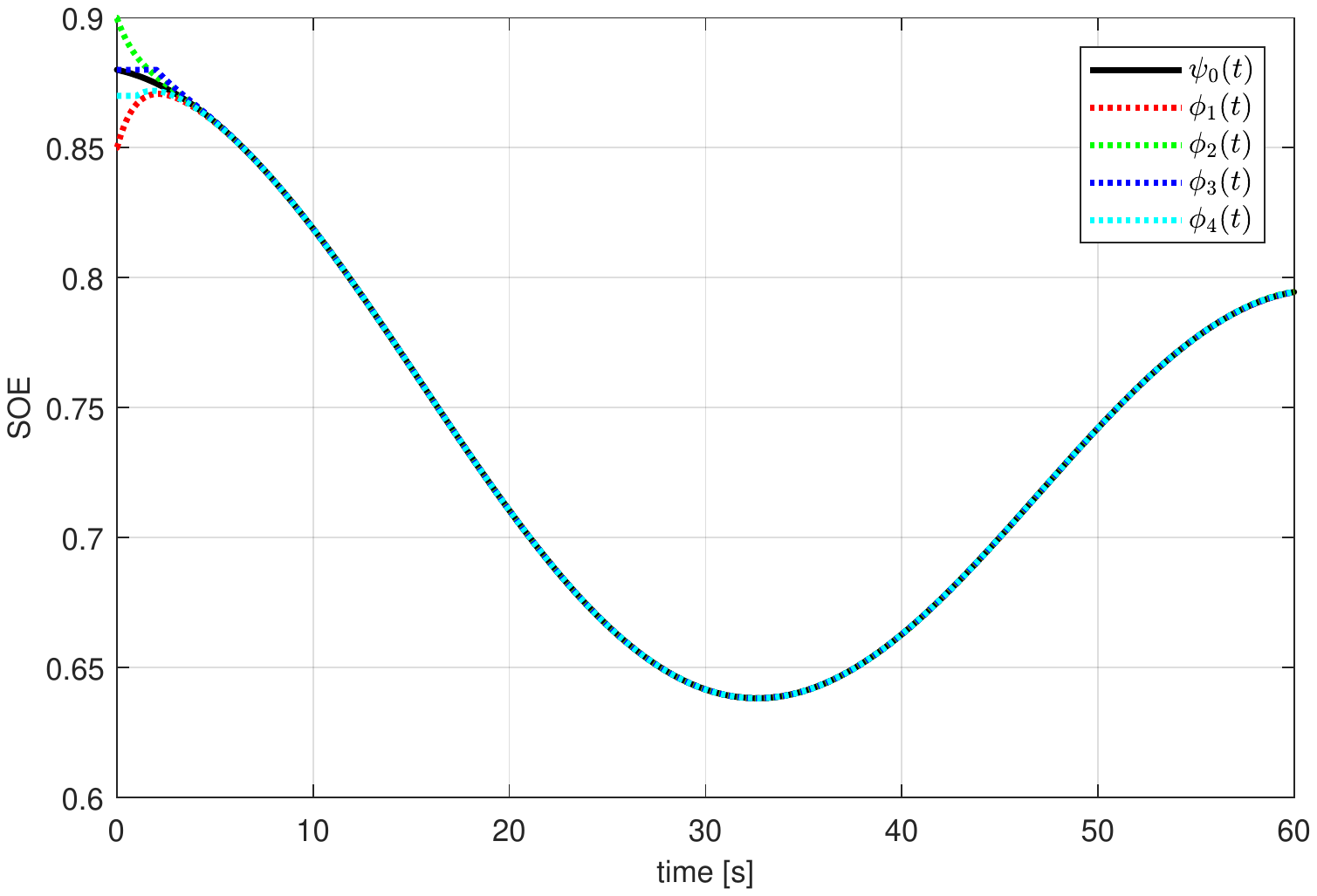}}
\caption{Performance on SOE balancing of the FESMS.}\label{performancesoe}
\end{center}
\end{figure}

\begin{figure}
\begin{center}
\scalebox{0.6}{\includegraphics[viewport=40 230 530 580]{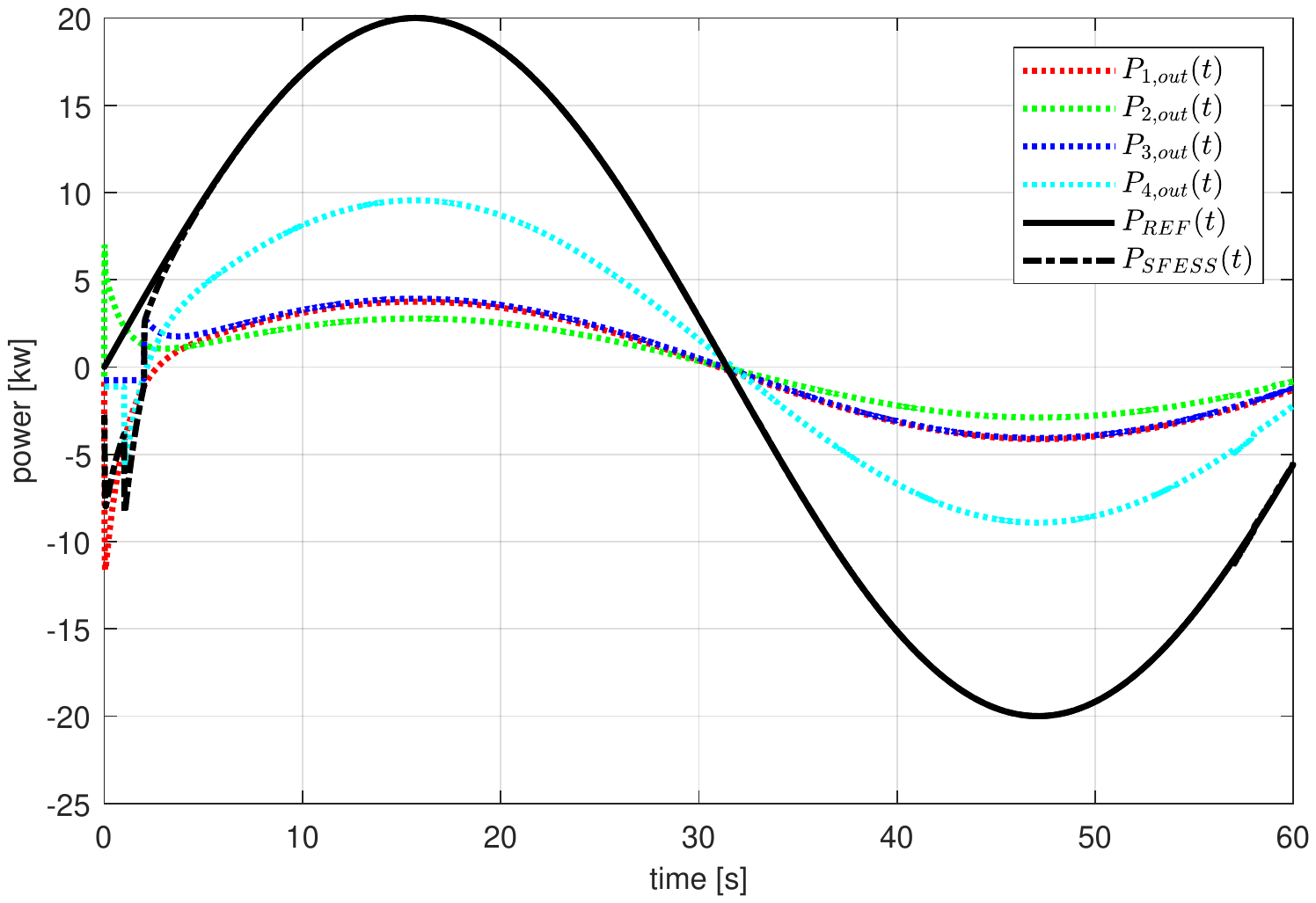}}
\caption{Performance on power tracking of the FESMS.}\label{performancepower}
\end{center}
\end{figure}

The control gains are selected to be $\mu_S=\mu_C=\mu_{\eta}=\mu_{\alpha}=\mu_{\beta}=\mu_{\psi}=100$, $\kappa=1$. The system initial values are given by $\psi_0(0)=0.88$,
$\phi_1(0)=\psi_1(0)=0.85$, $\phi_2(0)=\psi_2(0)=0.9$,
$\phi_3(0)=\psi_3(0)=0.88$, $\phi_4(0)=\psi_4(0)=0.87$,
and for $i=1,2,3,4$, $S_i(0)=0$, $C_i(0)=0$, $\eta_i(0)=0$, $\alpha_i(0)=0$, $\beta_i(0)=0$.
The system performance is shown by Figs. \ref{performancesoe} and \ref{performancepower}.
It can be seen that both SOE balancing and power tracking have been achieved.

\section{Conclusion}\label{seccon}

For a heterogenous FESMS, a distributed dual objective control problem aiming at
simultaneous reference power tracking and state-of-energy balancing has been considered.
It is first shown that a common SOE trajectory exists for all the flywheel systems
on which the dual control objectives can be achieved simultaneously. Then,
the distributed dual objective control
problem has been converted into a double layer distributed tracking problem by making use
of the common SOE trajectory, and is then solved by a distributed control law.

%\section*{Disclosure statement}
%
%No potential conflict of interest was reported by the authors.

\section*{Funding}

This work was supported by the National Natural
Science Foundation of China [grant number 61703167, 61803160] and
Guangdong Nature Science Foundation [grant number 2020A1515010810] and
Science and Technology Planning Project of Guangdong Province [grant number 2017A040405025] and
Guangzhou Science Research Programme [grant number 201904010242].

\section*{Notes on contributors}


\begin{thebibliography}{}
















\bibitem[Amirayr \& Pullen(2017)]{1zytie17}
Amirayr, M. E. and Pullen, K. R. (2017). A review of flywheel energy storage system technologies and their applications. \textit{Applied Sciences}, 7, 286.

\bibitem[Arami et. al.(2017)]{2zytie17}
Arami, A. K., Karami, H., Gharehpetian, G., and Hejazi, M. (2017). Review of flywheel energy storage systems structures and applications in power systems and microgrids. \textit{Renewable and Sustainable Energy Reviews}, 69:9-18.

\bibitem[Cai(2020)]{caiacess20}
Cai, H. (2020) Power tracking and state-of-energy balancing of an energy storage system by distributed control. \textit{IEEE Access}, 8:170261-170270.

\bibitem[Cai \& Hu(2016)]{caihutii16}
Cai, H., and Hu, G. (2016) Distributed control scheme for package-level state-of-charge balancing of grid-connected battery energy storage system. \textit{IEEE Transactions on Industrial Informatics}, 12(5):1919-1929.

%\bibitem[Cai \& Hu(2019)]{caihutcst19}
%Cai, H. and Hu, G. (2019). Distributed Robust Hierarchical Power Sharing Control of Grid-Connected Spatially Concentrated AC Microgrid. \textit{IEEE Transactions on Control Systems Technology}, vol. 27, no. 3, pp. 1012-1022.

\bibitem[Cao et. al.(2016)]{caotsg16}
Cao, Q., Song, Y., Guerrero, J. M., and Tian, S. (2016). Coordinated control for flywheel energy storage matrix systems for wind farm based on charging/discharging ratio consensus algorithms. \textit{IEEE Transactions on Smart Grid}, 7(3):1259-1267.

\bibitem[Chang et. al.(2015)]{4zytie14}
Chang, X., Li, Y., Zhang, W., Wang, N., and Xue, W. (2015). Active disturbance rejection control for a flywheel energy storage system. \textit{IEEE Transactions on Industrial Electronics}, 62(2):991-1001.



\bibitem[Dong \& Hu(2016)]{donghua16}
Dong, X. and Hu, G. (2016).
Time-varying formation control for general linear multi-agent systems with switching directed topologies,
\textit{Automatica},
73:47-55.



\bibitem[Dong et. al.(2017)]{dongzhoutie17}
Dong, X. Zhou, Y. Ren, Z. and Zhong, Y. (2017). Time-varying formation tracking for second-order multi-agent systems subjected to switching topologies with application to quadrotor formation flying. \textit{IEEE Transactions on Industrial Electronics}, 64(6):5014-5024.

%\bibitem[Elsayed \& Mohammed(2014)]{5zytie14}
%Elsayed, A. T. and Mohammed, O. A.(2014). Drstributed flywheel energy storage systems for mitigating the effects of pulsed loads. \textit{In 2014 IEEE PES General Meeting Conference Exposition},1-5.

\bibitem[Ghanaatian \& Lotfifard(2019)]{gltse19}
Ghanaatian, M.  and Lotfifard, S. (2019). Control of flywheel energy storage systems in the presence of uncertainties. \textit{IEEE Transactions on Sustainable Energy}, 10(1):36-45.

%\bibitem[Hockney et. al. (2003)]{7zytie03}
%Hockney, R.L., Lansberry, G.B., Davidkovich,V., Larkins,W.T., and Muchnik,E.(2003). Multiple flywheel energy storage systems. US Patent 6,614,132.

%\bibitem[Jin(2007)]{8zytie07}
%Jin,J.X.(2007). Hts energy storage techniques for use in distributed generation systems. \textit{Physica C: Superconductivity},460-462,1449-1450. Proceeding of the 8th International Conference on Materials and Mechanisms of Superconductivity and High Temperature Superconductors.

%\bibitem[Liu \& Jiang(2007)]{9zytie07}
%Liu, H and Jiang, J. (2007). Flywheel energy storage an upswing technology for energy sustainability.\textit{ Energy and Buildings},39(5),599-604.

\bibitem[Lai et. al.(2018)]{laitm18}
Lai, J. Song Y. and Du, X. (2018). Hierarchical coordinated control of flywheel energy storage matrix systems for wind farms. \textit{IEEE/ASME Transactions on Mechatronics}, 23(1):48-56.

\bibitem[Li et. al.(2017)]{litia17}
Li, C., Coelho, E. A. A., Dragicevic, T., Guerrero, J. M., and Vasquez, J. C. (2017). Multiagent-based distributed state of charge balancing control for distributed energy storage units in AC microgrids. \textit{IEEE Transactions on Industry Applications}, 53(3):2369-2381.

 \bibitem[Liu \& Huang(2019)]{lhijrnc19}
 Liu, T. and Huang, J. (2019). Leader-Following consensus with disturbance rejection for uncertain
Euler-Lagrange systems over switching networks. \textit{Int J Robust Nonlinear Control}. 29:6638-6656.

%\bibitem[Molina(2012)]{10zytie12}
%Molina,M.G.(2012). Distributed energy storage systems for applications in future smart grids. \textit{In 2012 Sixth IEEE/PES Transmission and Distribution:Latin America Conference and Exposition(T D-LA)},1-7.

\bibitem[Magdi et. al.(2016)]{magdiigss16}
Magdi S. Mahmoud, Mohamed Saif Ur Rahman and Fouad M. AL-Sunni (2016).
Networked control of microgrid system of systems, \textit{International Journal of Systems Science}, 47:11,
2607-2619.

\bibitem[Morstyn et. al.(2015)]{Morstyntps15}
Morstyn, T., Hredzak, B., and Agelidis, V. G. (2015). Distributed cooperative control of microgrid storage.
\textit{IEEE Transactions on Power Systems}, 30(5):2780-2789.

\bibitem[Mousavi et. al.(2017)]{11zytie17}
Mousavi, S., Faraji, F., Majazi, A., and Al-Haddad, K. (2017). A comprehensive review of flywheel energy storage system technology. \textit{Rennewable and Suatainable Energy Reviews}, 67, 477-490.

%\bibitem[Nguyen \& Tseng(2011)]{12zytie11}
%Nguyen,T.D., Tseng,K., Zhang,S., and Nguyen,H.T.(2011). A novel axial flux permanent-magnet machine for flywheel energy storage system: Design and analysis. \textit{IEEE Transaction on Industrial Electronics},58(9),3784-3794.

%\bibitem[Ren \& Cao(2011)]{13zytie11}
%Ren, W. and Cao, Y.(2011). \textit{Distributed Coordinations of Multi-agent Networks: Emergent Problems,Models, and Issues}. Springer,Springer-Verlag London.
%
%\bibitem[Su \& Huang(2012)]{14zytie12}
%Su, Y. and Huang, J.(2012). Cooperative output regulation of linear multi-agent systems. \textit{IEEE Transactions on Automatic Control},57(4),1062-1066.

\bibitem[Su \& Huang(2012)]{shtsmcb12}
Su, Y. and Huang, J. (2012). Cooperative output regulation with application to multi-agent consensus under switching network. \textit{IEEE Transactions on Systems, Man, and Cybernetics, Part B (Cybernetics)}, 42(3):864-875.

%\bibitem[Sun \& Dragicevic(2015)]{15zytie15}
%Sun,B., Dragicevic,T., Vasquez,J.C., and Guerrero,J.M.(2015). Distributed cooperative control of multi flywheel energy storage system for electrical vehicle fast charging stations. \textit{In 2015 17th European Conference on Power Eletronics and Applications(EPE'15 ECCE-Europe)},1-8.

\bibitem[Sun et. al.(2020)]{sunietcta20}
Sun, Y.  Hu J. and Liu, J. (2020). Periodic event-triggered control of flywheel energy storage matrix systems for wind farms. \textit{IET Control Theory \& Applications}, 14(11):1467-1477.

\bibitem[Vallejo et. al.(2014)]{Vallejoigss14}
Vallejo, D. Albusac, J. Glez-Morcillo, C. Castro-Schez J.J. and Jiménez L. (2014).
A multi-agent approach to intelligent monitoring in smart grids, \textit{International Journal of Systems
Science}, 45:4, 756-777.


\bibitem[Zhang \& Lewis(2018)]{zhangijss18}
Zhang, X., and Lewis, F. L. (2018). Cooperative output regulation of
heterogeneous multi-agent systems based on passivity, \textit{International Journal of Systems Science},
49:16, 3418-3430.


\bibitem[Zhang \& Yang(2017)]{zytie17}
Zhang, X. and Yang, J. (2017). A robust flywheel energy storage system discharge strategy for wide speed range operation. \textit{IEEE Transactions on Industrial Electronics}, 64(10):7862-7873.





\end{thebibliography}
\end{document}